\documentclass[aps,prd,superscriptaddress,preprintnumbers,nofootinbib]{revtex4}
\usepackage[latin9]{inputenc}
\usepackage{amsmath}
\usepackage{amssymb}
\usepackage{esint}
\usepackage[unicode=true,bookmarks=false, breaklinks=false,pdfborder={0 0 1},colorlinks=false]{hyperref}
\usepackage{breakurl}

\makeatletter
\@ifundefined{textcolor}{}
{%
 \definecolor{BLACK}{gray}{0}
 \definecolor{WHITE}{gray}{1}
 \definecolor{RED}{rgb}{1,0,0}
 \definecolor{GREEN}{rgb}{0,1,0}
 \definecolor{BLUE}{rgb}{0,0,1}
 \definecolor{CYAN}{cmyk}{1,0,0,0}
 \definecolor{MAGENTA}{cmyk}{0,1,0,0}
 \definecolor{YELLOW}{cmyk}{0,0,1,0}
}

\usepackage{amsfonts}
\usepackage{graphicx}
\usepackage{xcolor}
\usepackage{wrapfig}
\usepackage{enumerate}
\providecommand{\U}[1]{\protect\rule{.1in}{.1in}}
\tighten
\renewcommand{\d}{\partial}

\newcommand{\rd}{\mathcal}

\newcommand{\be}{\begin{eqnarray}}
\newcommand{\en}{\end{eqnarray}}
\newcommand{\badat}{\begin{alignedat}}
\newcommand{\eadat}{\end{alignedat}}
\newcommand{\bitm}{\begin{itemize}}
\newcommand{\eitm}{\end{itemize}}
\newcommand{\bmat}{\begin{pmatrix}}
\newcommand{\emat}{\end{pmatrix}}
\newcommand{\ba}{\begin{align}}
\newcommand{\bas}{\begin{align*}}
\newcommand{\ab}{\end{align}}
\newcommand{\bse}{\begin{subequations}}
\newcommand{\ese}{\end{subequations}}

\newcommand{\ep}{\epsilon}
\newcommand{\virg}{\hspace{1 mm}, \hspace{8 mm}}

\makeatother

\begin{document}

\title{Super-BMS$_{3}$ invariant boundary theory \\
 from three-dimensional flat supergravity}

\author{Glenn Barnich}
\email{gbarnich@ulb.ac.be}

\affiliation{Physique Théorique et Mathématique Université Libre de Bruxelles,
and International Solvay Institutes, Campus Plaine C.P. 231, B-1050
Bruxelles, Belgium.}

\author{Laura Donnay}
\email{ldonnay@ulb.ac.be}

\affiliation{Physique Théorique et Mathématique Université Libre de Bruxelles,
and International Solvay Institutes, Campus Plaine C.P. 231, B-1050
Bruxelles, Belgium.}

\author{Javier Matulich}

\email{matulich@cecs.cl}
\affiliation{Centro de Estudios Cient\'{i}ficos (CECs), Av. Arturo Prat 514, Valdivia,
Chile.}

\author{Ricardo Troncoso}

\email{troncoso@cecs.cl}
\affiliation{Centro de Estudios Cient\'{i}ficos (CECs), Av. Arturo Prat 514, Valdivia,
Chile.}

\preprint{CECS-PHY-15/09}
\begin{abstract}
The two-dimensional super-BMS$_{3}$ invariant theory dual to three-dimensional
asymptotically flat $\rd N=1$ supergravity is constructed. It is
described by a constrained or gauged chiral Wess-Zumino-Witten action
based on the super-Poincaré algebra in the Hamiltonian, respectively
the Lagrangian formulation, whose reduced phase space description
corresponds to a supersymmetric extension of flat Liouville theory. 
\end{abstract}
\maketitle

\section{Introduction}

A prime example of duality between a three-dimensional and a two-dimensional
theory is the relation between a Chern-Simons theory in the presence
of a boundary and the associated chiral Wess-Zumino-Witten (WZW) model:
on the classical level for instance, the variational principles are
equivalent as the latter is obtained from the former by solving
the constraints in the action \cite{Witten:1988hf,Moore:1989yh,Elitzur:1989nr}.

In the case of the Chern-Simons formulation of three-dimensional gravity
\cite{Achucarro:1987vz,Witten:1988hc}, the role of the boundary is
played by non trivial fall-off conditions for the gauge fields. For
anti-de Sitter or flat asymptotics, a suitable boundary term is required
in order to make solutions with the prescribed asymptotics true extrema
of the variational principle. Furthermore, the fall-off conditions
lead to additional constraints that correspond to fixing a subset
of the conserved currents of the WZW model \cite{Coussaert:1995zp,Henneaux:1999ib}.
The associated reduced phase space description is given by a Liouville
theory for negative cosmological constant and a suitable limit thereof
in the flat case \cite{Barnich:2012rz,Barnich:2013yka}.

In this paper, we apply the construction to three-dimensional asymptotically
flat $\rd N=1$ supergravity, whose algebra of surface charges has
been shown to realize the centrally extended super-BMS$_{3}$ algebra
\cite{Barnich:2014cwa}. The non-vanishing Poisson brackets read 
\begin{equation}
\begin{split}i\{\mathcal{J}_{m},\mathcal{J}_{n}\} & =(m-n)\mathcal{J}_{m+n}+\frac{c_{1}}{12}m^{3}\delta_{m+n,0}\,,\\
i\{\mathcal{J}_{m},\mathcal{P}_{n}\} & =(m-n)\mathcal{P}_{m+n}+\frac{c_{2}}{12}m^{3}\delta_{m+n,0}\,,\\
i\{\mathcal{J}_{m},\mathcal{Q}_{n}\} & =\left(\frac{m}{2}-n\right)\mathcal{Q}_{m+n}\,,\\
\{\mathcal{Q}_{m},\mathcal{Q}_{n}\} & =\mathcal{P}_{m+n}+\frac{c_{2}}{6}m^{2}\delta_{m+n,0}\,,
\end{split}
\label{eq:Super-BMS-simplest}
\end{equation}
where the fermionic generators $\mathcal{Q}_{m}$ are labeled by (half-)integers
in the case of (anti)periodic boundary conditions for the gravitino,
and the central charges are given by 
\begin{equation}
c_{1}=\mu\frac{3}{G}\quad,\quad c_{2}=\frac{3}{G}\,.\label{eq:1}
\end{equation}
Here, $G$ and $\mu$ stand for the Newton constant and the coupling
of the Lorentz-Chern-Simons form, respectively.

The resulting two-dimensional field theory admits a global super-BMS$_{3}$
invariance. By construction, the associated algebra of Noether charges
realizes \eqref{eq:Super-BMS-simplest} with the same values of the
central charges. We provide three equivalent descriptions of this
theory: (i) a Hamiltonian description in terms of a constrained chiral
WZW theory based on the three-dimensional super-Poincaré algebra,
(ii) a Lagrangian formulation in terms of a gauged chiral WZW theory
and (iii) a reduced phase space description that corresponds to a 
supersymmetric extension of flat Liouville theory.

Besides the extension to the supersymmetric case, previous results
in the purely bosonic sector are also generalized. This is due to
the inclusion of parity-odd terms in the action, which suitably modifies
the Poincaré current subalgebra, and consequently, turns on the additional
central charge $c_{1}$ in \eqref{eq:Super-BMS-simplest}.

\section{Brief review of (minimal) $\mathcal{N}=1$ flat supergravity in 3D}

\label{Standard N=00003D00003D00003D1 supergravity}

As in the case of pure gravity, minimal $\mathcal{N}=1$ supergravity
in three dimensions \cite{Deser:1982sw,Deser:1984aa,Marcus:1983hb}
with vanishing cosmological constant admits a Chern-Simons formulation
\cite{Achucarro:1989gm}. Different extensions of this theory have
been developed in e.g., \cite{Nishino:1991sr,Howe:1995zm,Banados:1996hi,Giacomini:2006dr,Gupta:2007th,Andringa,Bergshoeff:2010ui,Bergshoeff:2013ida,Andringa:2013mma,Fierro:2014lka,Alkac:2014hwa,Fuentealba:2015jma,Concha:2015woa,Bergshoeff:2015ija}.
Hereafter we consider the most general supergravity theory with $\mathcal{N}=1$
that is compatible with asymptotically flat boundary conditions, and
leads to first order field equations for the dreibein, the spin connection
and the gravitino \cite{Giacomini:2006dr}(see also \cite{Barnich:2014cwa,Fuentealba:2015wza}).
The standard minimal $\mathcal{N}=1$ supergravity theory is recovered
for a particular choice of the couplings (see below). The gauge field
$A=A_{\mu}dx^{\mu}$ is given by 
\begin{equation}
A=e^{a}P_{a}+\hat{\omega}^{a}J_{a}+\psi^{\alpha}Q_{\alpha}\,,
\end{equation}
where $e^{a}$, $\omega^{a}$ and $\psi^{\alpha}$ stand for the dreibein,
the dualized spin connection $\omega_{a}=\frac{1}{2}\epsilon_{abc}\omega^{bc}$,
and the (Majorana) gravitino, respectively; while $\hat{\omega}^{a}:=\omega^{a}+\gamma e^{a}$
and the set $\{P_{a},J_{a},Q_{\alpha}\}$ spans the super-Poincaré
algebra, 
\begin{equation}
\begin{alignedat}{2}
&\lbrack J_{a},J_{b}]=\epsilon_{abc}J^{c}\quad,\quad\lbrack J_{a},P_{b}]=\epsilon_{abc}P^{c}\quad,\quad\lbrack P_{a},P_{b}]=0\,, \\
&\lbrack J_{a},Q_{\alpha}]=\frac{1}{2}\left(\Gamma_{a}\right)_{\hspace{2mm}\alpha}^{\beta}Q_{\beta}\quad,\quad\lbrack P_{a},Q_{\alpha}]=0\quad,\quad\{Q_{\alpha},Q_{\beta}\}=-\frac{1}{2}\left(C\Gamma^{a}\right)_{\alpha\beta}P_{a}\,,\label{Super Poincare}
\end{alignedat}
\end{equation}
where $C$ is the charge conjugation matrix (see Appendix \ref{conventions}
for conventions). In these terms, the action reads 
\begin{equation}
I[A]=\frac{k}{4\pi}\int\langle A,dA+\frac{2}{3}A^{2}\rangle\,,\label{ICS}
\end{equation}
where the bracket $\langle\cdot,\cdot\rangle$ stands for an invariant
nondegenerate bilinear form, whose only nonvanishing components are
given by 
\begin{equation}
\langle P_{a},J_{b}\rangle=\eta_{ab}\quad,\quad\langle
J_{a},J_{b}\rangle=
\mu\eta_{ab}\quad,\quad\langle
Q_{\alpha},Q_{\beta}\rangle=C_{\alpha\beta}\,,\label{Bracket}
\end{equation}
and the level is related to the Newton constant according to $k=\frac{1}{4G}$.
Hence, up to a boundary term, the action reduces to 
\begin{equation}
I_{(\mu,\gamma)}=\frac{k}{4\pi}\int2\hat{R}^{a}e_{a}+\mu
L(\hat{\omega})-
\bar\psi_{\alpha}\hat{D}\psi^{\alpha}\,,
\end{equation}
where $\bar{\psi}_{\alpha}=C_{\alpha\beta}\psi^{\beta}$ is the Majorana
conjugate, and with respect to the connection $\hat{\omega^{a}}$,
the curvature two-form and the covariant derivative of the gravitino
are defined as 
\begin{equation}
\hat{R}^{a}=d\hat{\omega}^{a}+\frac{1}{2}\epsilon^{abc}\hat{\omega}_{b}
\hat{\omega}_{c}\quad,\quad\hat{D}\psi=d\psi+\frac{1}{2}\hat{\omega}^{a}\Gamma_{a}\psi\,,
\end{equation}
respectively, while
$L(\hat{\omega})=\hat{\omega}^{a}d\hat{\omega}_{a}+
\frac{1}{3}\epsilon_{abc}\hat{\omega}^{a}\hat{\omega}^{b}\hat{\omega}^{c}$
is the corresponding Lorentz-Chern-Simons form.

By construction the action is invariant, up to a surface term, under
the following local supersymmetry transformations 
\begin{equation}
\delta e^{a}=-\frac{1}{2}\bar{\epsilon}\Gamma^{a}\psi\quad,\quad\delta\omega^{a}=\frac{1}{2}\gamma\bar{\epsilon}\Gamma^{a}\psi\quad,\quad\delta\psi=D\epsilon+\frac{1}{2}\gamma e^{a}\Gamma_{a}\epsilon\,,\label{susy reloaded}
\end{equation}
where $D\epsilon=d\epsilon+\frac{1}{2}\omega^{a}\Gamma_{a}\epsilon$
is the standard Lorentz covariant derivative of a spinor. The field
equations $F=dA+A^{2}=0$, whose general solution is locally given
by $A=G^{-1}dG$, decompose as

\begin{equation}
R^{a}=\frac{1}{2}\gamma^{2}\epsilon^{abc}e_{b}e_{c}+\frac{1}{4}\gamma\bar{\psi}\Gamma^{a}\psi\quad,\quad T^{a}=-\gamma\epsilon^{abc}e_{b}e_{c}-\frac{1}{4}\bar{\psi}\Gamma^{a}\psi\quad,\quad D\psi=-\frac{1}{2}\gamma e^{a}\Gamma_{a}\psi\,,\label{Feqs-1}
\end{equation}
where $R^{a}$, and $T^{a}=de^{a}+\epsilon^{abc}\omega_{b}e_{c}$
stand for the curvature and torsion two-forms, respectively.

Defining $\hat{\omega}=\frac{1}{2}\hat{\omega}^{a}\Gamma_{a}$, $e=\frac{1}{2}e^{a}\Gamma_{a}$
and contracting the first two equations in \eqref{Feqs-1} with $\frac{1}{2}\Gamma_{a}$
gives the matrix form $d\hat{\omega}+\hat{\omega}^{2}=0$, $de+[\hat{\omega},e]=-\frac{1}{4}\psi\bar{\psi}$,
so that the decomposition of the general (local) solution is 
\begin{equation}
\hat{\omega}=\Lambda^{-1}d\Lambda\quad,\quad\psi=\Lambda^{-1}d\eta\quad,\quad e=\Lambda^{-1}(-\frac{1}{4}\eta d\bar{\eta}-\frac{1}{8}d\bar{\eta}\eta\mathbf{1}+db)\Lambda\,,\label{eq:7}
\end{equation}
where $\Lambda$ is an $\mathrm{SL}(2,\mathbf{\mathbb{R}})$ group
element, $\eta$ a Grassmann-valued spinor and $b$ a traceless $2\times2$
matrix.

The asymptotic conditions proposed in \cite{Barnich:2014cwa} imply
that the gauge field is of the form 
\begin{equation}
A=h^{-1}ah+h^{-1}dh\,,\label{A}
\end{equation}
where the radial dependence is completely captured by the group element
$h=e^{-rP_{0}}$, while 
\begin{equation}
a=\left(\frac{\mathcal{M}}{2}du+\frac{\mathcal{N}}{2}d\phi\right)P_{0}+du\, P_{1}+\frac{\mathcal{M}}{2}d\phi J_{0}+d\phi\, J_{1}+\frac{\psi}{2^{1/4}}d\phi\, Q_{+}\,,\label{flat-connection}
\end{equation}
with functions $\mathcal{M}$, $\mathcal{N}$, and the Grassmann-valued
spinor component $\psi$ that depend on the remaining coordinates
$u$, $\phi$.

The standard supergravity theory with $\mathcal{N}=1$ with its asymptotically
flat behaviour is then recovered for $\mu=\gamma=0$. It is also worth
pointing out that the fall-off conditions \eqref{flat-connection}
can be generalized, along the lines of \cite{Henneaux_ChP}, so as
to include a generic choice of chemical potentials \cite{Fuentealba:2015wza}.

\section{Chiral constrained super-Poincaré WZW theory}

\label{sec:chir-super-poinc}

\subsection{Solving the constraints in the action}

\label{sec:solv-constr-acti}

Up to boundary terms and an overall sign which we change for later
convenience, the Hamiltonian form of the Chern-Simons action \eqref{ICS}
is given by 
\begin{equation}
I_{H}[A]=-\frac{k}{4\pi}\int\langle\tilde{A},du\dot{\tilde{A}}\rangle+2\langle duA_{u},\tilde{d}\tilde{A}+\tilde{A}^{2}\rangle\,,\label{eq:2}
\end{equation}
where $A=duA_{u}+\tilde{A}$.

One of the advantages of the gauge choice in \eqref{A}, for which
the dependence in the radial coordinate is completely absorbed by
the group element $h$, is that the boundary can be assumed to be
unique and located at an arbitrary fixed value of $r=r_{0}$. Hence,
the boundary generically describes a two-dimensional timelike surface
with the topology of a cylinder ($\mathbb{R}\times S^{1}$). We will
also discard all holonomy terms. As a consequence, the resulting action
principle at the boundary only captures the asymptotic symmetries
of the original gravitational theory. Note also that positive orientation
in the bulk is taken as $dud\phi dr$.

The boundary term in the variation of the Hamiltonian action is given
by $-\frac{k}{2\pi}du\tilde{d}\langle A_{u},\delta\tilde{A}\rangle$.
Thus, by virtue of the boundary conditions \eqref{flat-connection},
the components of the gauge field at the boundary fulfill 
\begin{eqnarray}
\omega_{\phi}^{a}=e_{u}^{a}\quad,\quad\omega_{u}^{a}=0\quad,\quad\psi_{u}^{+}=0=\psi_{u}^{-}\,,
\end{eqnarray}
so that the boundary term becomes integrable. Consequently, the improved
action principle that has a true extremum when the equations of motion
are satisfied is given by 
\begin{equation}
I_{I}[A]=I_{H}[A]-\frac{k}{4\pi}\int dud\phi\,\omega_{\phi}^{a}\omega_{a\phi}|^{r=r_{0}}\,.
\end{equation}
In this action principle $A_{u}$ are Lagrange multipliers, whose
associated constraints are locally solved by $\tilde{A}=G^{-1}\tilde{d}G$
for some group element $G(u,r,\phi)$. Solving the constraints in
the action yields 
\begin{equation}
I=\frac{k}{4\pi}\left(\int dud\phi\,\big[\langle G^{-1}\d_{\phi}G,G^{-1}\d_{u}G\rangle-\omega_{\phi}^{a}\omega_{a\phi}\big]^{r=r_{0}}+\Gamma[G]\right)\,,\label{eq:6}
\end{equation}
where 
\begin{equation}
\Gamma[G]=\frac{1}{3}\int\langle G^{-1}dG,(G^{-1}dG)^{2}\rangle\,.\label{eq:4}
\end{equation}
Equivalently, in terms of the gauge field components, the action can
be conveniently written as 
\begin{equation}
I=\frac{k}{4\pi}\left(\int dud\phi\,\big[\omega_{\phi}^{a}e_{au}+e_{\phi}^{a}\omega_{au}-\omega_{\phi}^{a}\omega_{a\phi}+\mu\omega_{\phi}^{a}\omega_{au}-\bar{\psi}_{u}\psi_{\phi}\big]^{r=r_{0}}+\Gamma[G]\right)\,,\label{eq:3}
\end{equation}
with 
\begin{equation}
\Gamma[G]=\frac{1}{6}\int(3\epsilon_{abc}e^{a}\omega^{b}\omega^{c}+\mu\epsilon_{abc}\omega^{a}\omega^{b}\omega^{c}-\frac{3}{2}\omega^{a}(C\Gamma_{a})_{\alpha\beta}\psi^{\alpha}\psi^{\beta})\,,
\end{equation}
and the understanding that $A_{\mu}=G^{-1}\d_{\mu}G$. Decomposing
this connection according to eq. \eqref{eq:7} allows one to rewrite
this expression in terms of a $2$ by $2$ matrix trace, so that integrating
by parts the first term in $\Gamma[G]$ gives 
\begin{equation}
I=\frac{k}{2\pi}\int dud\phi\,{\rm Tr}\big[2\dot{\Lambda}\Lambda^{-1}(-\frac{\eta\bar{\eta}'}{4}+b')-(\Lambda'\Lambda^{-1})^{2}+\mu\Lambda'\Lambda^{-1}\dot{\Lambda}\Lambda^{-1}+\frac{\eta'\dot{\bar{\eta}}}{2}\big]^{r=r_{0}}+\frac{\mu}{3}\int{\rm Tr}(d\Lambda\Lambda^{-1})^{3}\,.\label{eq:9}
\end{equation}
Furthermore, the boundary conditions \eqref{A}, \eqref{flat-connection}
imply that $\partial_{\phi}A_{r}=0$, and hence $G=g(u,\phi)h(u,r)$.
More precisely, since in the asymptotic region $h=e^{-rP_{0}}$, one obtains in
particular that $\dot{h}(u,r_{0})=0$. The decomposition in
\eqref{eq:7} is then refined as
\begin{eqnarray}
\badat{3} & \Lambda=\lambda(u,\phi)\,\varsigma(u,r)\,,\\
 & \eta=\nu(u,\phi)+\lambda\varrho(u,r)\,,\\
 & b=\alpha(u,\phi)+\frac{1}{4}\nu\bar{\varrho}\lambda^{-1}+\frac{1}{8}\bar{\varrho}\lambda^{-1}\nu\mathbf{1}+\lambda\beta(u,r)\lambda^{-1}\,,\eadat
\end{eqnarray}
where $\dot{\varsigma}(u,r_{0})=\dot{\varrho}(u,r_{0})=\dot{\beta}(u,r_{0})=0$.
Therefore, up to a total derivative in $u$ and $\phi$, one finds that
the action reduces to that of a chiral super-Poincaré Wess-Zumino-Witten theory,
\begin{multline}
I[\lambda,\alpha,\nu]=\frac{k}{2\pi}\int dud\phi\,{\rm Tr}\big[2\dot{\lambda}\lambda^{-1}\alpha'-(\lambda'\lambda^{-1})^{2}+\mu\lambda'\lambda^{-1}\dot{\lambda}\lambda^{-1}+\frac{1}{2}\nu'\dot{\bar{\nu}}-\frac{1}{2}\dot{\lambda}\lambda^{-1}\nu\bar{\nu}'\big]+\frac{\mu}{3}\int{\rm Tr}(d\Lambda\Lambda^{-1})^{3}\,.\\
\label{eq:SuperWZW}
\end{multline}
The field equations are then obtained by varying \eqref{eq:SuperWZW}
with respect to $\alpha,\nu,\lambda$, which gives
\begin{eqnarray}
\badat{3} & (\dot{\lambda}\lambda^{-1})'=0\,,\label{eq:5}\\
 & D_{u}^{-\dot{\lambda}\lambda^{-1}}\nu'=0\,,\\
 & D_{u}^{-\dot{\lambda}\lambda^{-1}}\alpha'+(\mu\partial_{u}-\partial_{\phi})(\lambda'\lambda^{-1})-\frac{1}{4}\dot{\nu}\bar{\nu}'-\frac{1}{8}\bar{\nu}'\dot{\nu}\mathbf{1}+\frac{1}{4}\dot{\lambda}\lambda^{-1}\nu\bar{\nu}'+\frac{1}{8}\bar{\nu}'\dot{\lambda}\lambda^{-1}\nu\mathbf{1}=0\,,\eadat
\end{eqnarray}
respectively. The general solution of these equations is given by
\begin{eqnarray}
\badat{3} & \lambda=\tau(u)\kappa(\phi)\,,\label{eq:11}\\
 & \nu=\tau(\zeta_{1}(u)+\zeta_{2}(\phi))\,,\\
 & \alpha=\tau\Big(a(\phi)+\delta(u)+u\kappa'\kappa^{-1}-\mu[\ln\tau,\ln\kappa]+\frac{1}{4}\zeta_{1}\bar{\zeta_{2}}+\frac{1}{8}\bar{\zeta_{2}}\zeta_{1}\mathbf{1}\Big)\tau^{-1}\,.\eadat
\end{eqnarray}

\subsection{Symmetries of the chiral WZW model}

By using the Polyakov-Wiegmann identities, the action \eqref{eq:SuperWZW}
can be shown to be invariant under the gauge transformations 
\begin{equation}
\lambda\to\Xi(u)\lambda\quad,\quad\nu\to\Xi\nu\quad,\quad\alpha\to\Xi\alpha\Xi^{-1}\,.\label{eq:12}
\end{equation}
Moreover, it is also invariant under the following global symmetries
\begin{align}
\lambda\to\lambda & \quad, & \nu\to\nu & \quad, & \alpha\to\alpha+\lambda\Sigma(\phi)\lambda^{-1}\,,\nonumber \\
\lambda\to\lambda\Theta^{-1}(\phi) & \quad, & \nu\to\nu & \quad, & \alpha\to\alpha-u\lambda\Theta^{-1}\Theta'\lambda^{-1}\,,\label{eq:13}\\
\lambda\to\lambda & \quad, & \nu\to\nu+\lambda\Upsilon(\phi) & \quad, & \alpha\to\alpha+\frac{1}{4}\nu\bar{\Upsilon}\lambda^{-1}+\frac{1}{8}\bar{\Upsilon}\lambda^{-1}\nu\mathbf{1\,},\nonumber 
\end{align}
whose associated infinitesimal transformations read 
\begin{align}
\delta_{\sigma}\lambda=0 & \quad, & \delta_{\sigma}\nu=0 & \quad, & \delta_{\sigma}\alpha=\lambda\sigma(\phi)\lambda^{-1}\,,\nonumber \\
\delta_{\vartheta}\lambda=-\lambda\vartheta(\phi) & \quad, & \delta_{\vartheta}\nu=0 & \quad, & \delta_{\vartheta}\alpha=-u\lambda\vartheta'\lambda^{-1}\,,\label{eq:infin_global}\\
\delta_{\gamma}\lambda=0 & \quad, & \delta_{\gamma}\nu=\lambda\gamma(\phi) & \quad, & \delta_{\gamma}\alpha=\frac{1}{4}\nu\bar{\gamma}\lambda^{-1}+\frac{1}{8}\bar{\gamma}\lambda^{-1}\nu\mathbf{1}\,.\nonumber 
\end{align}
The Noether currents associated to a global symmetry, whose parameters
are collectively denoted by $X_{1}$, generically read $J_{X_{1}}^{\mu}=-k_{X_{1}}^{\mu}+\frac{\partial{\cal L}}{\partial_{\mu}\phi^{i}}\delta_{X_{1}}\phi^{i}$,
with $\delta_{X_{1}}{\cal L}=\partial_{\mu}k_{X_{1}}^{\mu}$. Hence,
in the case of global symmetries spanned by \eqref{eq:infin_global},
the corresponding currents are given by $J_{\sigma}^{\mu}=2\delta_{0}^{\mu}{\rm Tr}[\sigma P]$,
$J_{\vartheta}^{\mu}=2\delta_{0}^{\mu}{\rm Tr}[\vartheta J]$, $J_{\gamma}^{\mu}=2\delta_{0}^{\mu}{\rm Tr}[\gamma Q]$,
where 
\begin{eqnarray}
\badat{3} & P=\frac{k}{2\pi}\lambda^{-1}\lambda'\,,\label{eq:15}\\
 & J=-\frac{k}{2\pi}[\lambda^{-1}\alpha'\lambda-u(\lambda^{-1}\lambda')'+\mu\lambda^{-1}\lambda'-\frac{1}{4}\lambda^{-1}\nu\bar{\nu}'\lambda-\frac{1}{8}\bar{\nu}'\nu\mathbf{1}]\,,\\
 & Q=\frac{k}{4\pi}\bar{\nu}'\lambda\,.\eadat
\end{eqnarray}

For the Noether $n-1$-forms $j_{X_{1}}=J_{X_{1}}^{\mu}(d^{n-1}x)_{\mu}$,
the current algebra can then be worked out by applying a subsequent
symmetry transformation $\delta_{X_{2}}$, so that 
\begin{equation}
\delta_{X_{2}}j_{X_{1}}=j_{[X_{1},X_{2}]}+K_{X_{1},X_{2}}+{\rm ``trivial"\,},\label{eq:16}
\end{equation}
where $[\delta_{X_{1}},\delta_{X_{2}}]=\delta_{[X_{2},X_{1}]}$, and
$K_{X_{1},X_{2}}$ denotes a possible field dependent central extension,
and ``trivial'' stands for exact $n-1$ forms plus terms that vanish
on-shell. Furthermore, general results guarantee that, in the Hamiltonian
formalism, this computation corresponds to the Dirac bracket algebra
of the canonical generators of the symmetries, i. e., $\delta_{X_{2}}J_{X_{1}}^{0}=\{J_{X_{1}}^{0},J_{X_{2}}^{0}\}^{*}$,
see e.g.~\cite{Henneaux:1992ig,Barnich:1996mr,Barnich:2007bf,Barnich:2013axa}.
Once applied to the components of the currents, given by 
\begin{equation}
P_{a}(\phi)={\rm Tr}[\Gamma_{a}P]\quad,\quad J_{a}(\phi)={\rm Tr}[\Gamma_{a}J]\quad,\quad Q_{\alpha}(\phi)=-\frac{k}{2\pi}\bar{\nu}'_{\beta}\lambda_{\alpha}^{\beta}\,,\label{eq:17}
\end{equation}
this yields 
\begin{eqnarray}
\badat{4} & \{P_{a}(\phi),P_{b}(\phi')\}^{*}=0\,,\\
 & \{J_{a}(\phi),J_{b}(\phi')\}^{*}=\epsilon_{abc}J^{c}\delta(\phi-\phi')-\mu\frac{k}{2\pi}\eta_{ab}\partial_{\phi}\delta(\phi-\phi')\,,\\
 & \{J_{a}(\phi),P_{b}(\phi')\}^{*}=\epsilon_{abc}P^{c}\delta(\phi-\phi')-\frac{k}{2\pi}\eta_{ab}\partial_{\phi}\delta(\phi-\phi')\,,\\
 & \{P_{a}(\phi),Q_{\alpha}(\phi')\}^{*}=0\,,\\
 & \{J_{a}(\phi),Q_{\alpha}(\phi')\}^{*}=\frac{1}{2}(Q\Gamma_{a})_{\alpha}\delta(\phi-\phi')\,,\\
 & \{Q_{\alpha}(\phi),Q_{\beta}(\phi')\}^{*}=-\frac{1}{2}(C\Gamma^{a})_{\alpha\beta}P_{a}\delta(\phi-\phi')-\frac{k}{2\pi}C_{\alpha\beta}\partial_{\phi}\delta(\phi-\phi')\,,\eadat\label{eq:6a}
\end{eqnarray}
which is the affine extension of the super-Poincaré algebra \eqref{Super Poincare}.

\subsection{Super-BMS$_{3}$ algebra from a modified Sugawara construction}

\label{sec:modif-sugaw-constr}

In order to recover the super-BMS$_{3}$ algebra \eqref{eq:Super-BMS-simplest}
from the affine extension of the super-Poincaré algebra in \eqref{eq:6a},
it can be seen that the standard Sugawara construction has to be slightly
improved. Indeed, let us consider bilinears made out of the currents
components $P_{a}$, $J_{a}$, $Q_{\alpha}$, given by 
\begin{equation}
{\cal H}=\frac{\pi}{k}P^{a}P_{a}\quad,\quad{\cal P}=-\frac{2\pi}{k}J^{a}P_{a}+\mu{\cal H}+\frac{\pi}{k}Q_{\alpha}C^{\alpha\beta}Q_{\beta}\quad,\quad\mathcal{G}=2^{3/4}\frac{\pi}{k}\left(\, P_{2}\, Q_{+}+\,\sqrt{2}\, P_{0}\, Q_{-}\right)\,,\label{eq:19-1}
\end{equation}
for which the current algebra \eqref{eq:6a} implies 
\begin{align}
\{{\cal H}(\phi),P_{a}(\phi')\}^{*}=0 & \quad, & \{{\cal P}(\phi),P_{a}(\phi')\}^{*}=P_{a}(\phi)\delta'(\phi-\phi')\,,\nonumber \\
\{{\cal H}(\phi),J_{a}(\phi')\}^{*}=-P_{a}(\phi)\delta'(\phi-\phi') & \quad, & \{{\cal P}(\phi),J_{a}(\phi')\}^{*}=J_{a}(\phi)\delta'(\phi-\phi')\,,\label{eq:20}\\
\{{\cal H}(\phi),{Q}_{\alpha}(\phi')\}^{*}=0 & \quad, & \{{\cal P}(\phi),{Q}_{\alpha}(\phi')\}^{*}={Q}_{\alpha}(\phi)\delta'(\phi-\phi')\,,\nonumber 
\end{align}
\begin{eqnarray}
 &  & \{\mathcal{G}(\phi),P_{a}(\phi')\}^{*}=0\,,\nonumber \\
 &  & \{\mathcal{G}(\phi),J_{a}(\phi')\}^{*}=-\frac{\pi}{2^{1/4}k}(\epsilon_{abc}(Q\Gamma^{b})_{+}P^{c}+P_{a}Q_{+})\delta(\phi-\phi')-\delta'(\phi-\phi')\frac{1}{2^{1/4}}(Q\Gamma_{a})_{+}(\phi')\,,\\
 &  & \{\mathcal{G}(\phi'),Q_{\alpha}(\phi')\}^{*}=-\frac{1}{2^{1/4}}{\cal H}C_{\alpha+}\delta(\phi-\phi')+\delta'(\phi-\phi')\frac{1}{2^{1/4}}(C\Gamma_{a})_{\alpha+}P^{a}(\phi)\,.\nonumber 
\end{eqnarray}
When expressed in terms of modes, the algebra of the generators ${\cal
  H}$, ${\cal P}$ corresponds to the pure BMS$_{3}$ algebra without
central extensions, i.e., the bosonic part of
\eqref{eq:Super-BMS-simplest} with $c_1=0=c_2$. This does however not
hold for the mode expansion of the full set ${\cal H}$, ${\cal P}$,
$\mathcal{G}$ whose algebra disagrees with the non-centrally extended
super BMS$_{3}$ algebra given in \eqref{eq:Super-BMS-simplest}. It
reflects the fact that the non-constrained super-WZW model
\eqref{eq:SuperWZW} is invariant under global BMS$_{3}$
transformations, but not under the full super-BMS$_{3}$ symmetries, in
the sense that there are no (obvious) superpartners to $\rd H$, $\rd
P$ that would close with them according to the (non centrally
extended) super-BMS algebra (see \cite{Henneaux:1999ib} for an analogous
discussion in the case of the superconformal algebra).

According to the fall-off of the gauge field in
\eqref{flat-connection}, the remaining boundary conditions that have
to be taken into account imply that $[\lambda^{-1}\lambda']^{1}=1$,
$[\lambda^{-1}\nu']^{-}=0$,
$[\lambda^{-1}(-\frac{1}{4}\nu\bar{\nu}'-\frac{1}{8}\bar{\nu}'\nu\mathbf{1}+\alpha')\lambda]^{1}=0$.
In terms of the currents, this amounts to imposing the following first
class constraints
\begin{eqnarray}
P_{0}=\frac{k}{2\pi}\quad,\quad J_{0}=-\frac{\mu k}{2\pi}\quad,\quad Q_{+}=0\,.\label{constraints}
\end{eqnarray}
The super-BMS$_{3}$ invariance of our model with the correct values
of the central charges is recovered only once the constraints \eqref{constraints}
are imposed. The generators of super-BMS$_{3}$ symmetry in the constrained
theory are given by 
\begin{eqnarray}
\badat{3} & \tilde{\rd H}=\rd H+\partial_{\phi}P_{2},\\
 & \tilde{\rd P}=\rd P-\partial_{\phi}J_{2},\label{eq:super_BMS_generators}\\
 & \tilde{\mathcal{G}}=\mathcal{G}+2^{3/4}\partial_{\phi}Q_{+}(\phi),\eadat
\end{eqnarray}
which are representatives that commute with the first class
constraints \eqref{constraints}, on the surface defined by these
constraints. Furthermore, on this surface, the Dirac brackets of the
generators are given by
\begin{eqnarray}
\badat{4}\{{\cal \tilde{H}}(\phi),{\cal \tilde{H}}(\phi')\}^{*} & =0\,,\\
\{{\cal \tilde{H}}(\phi),{\cal \tilde{P}}(\phi')\}^{*} & =({\cal \tilde{H}}(\phi)+{\cal \tilde{H}}(\phi'))\partial_{\phi}\delta(\phi-\phi')-\frac{k}{2\pi}\partial_{\phi}^{3}\delta(\phi-\phi')\,,\\
\{{\cal \tilde{P}}(\phi),{\cal \tilde{P}}(\phi')\}^{*} & =({\cal \tilde{P}}(\phi)+{\cal \tilde{P}}(\phi'))\partial_{\phi}\delta(\phi-\phi')-\frac{\mu k}{2\pi}\partial_{\phi}^{3}\delta(\phi-\phi')\,,\\
\{{\cal \tilde{H}}(\phi),{\cal \tilde{\mathcal{G}}}(\phi')\}^{*} & =0\,,\\
\{{\cal \tilde{P}}(\phi),{\cal \tilde{\mathcal{G}}}(\phi')\}^{*} & =(\tilde{\mathcal{G}}(\phi)+\frac{1}{2}{\cal \tilde{\mathcal{G}}}(\phi'))\partial_{\phi}\delta(\phi-\phi')\,,\\
\{\tilde{\mathcal{G}}(\phi),\tilde{\mathcal{G}}(\phi')\}^{*} & ={\cal \tilde{H}}(\phi)\delta(\phi-\phi')-\frac{k}{\pi}\partial_{\phi}^{2}\delta(\phi-\phi')\,,\eadat
\end{eqnarray}
so that, once expanded in modes according to 
\[
{\cal P}_{m}=\int_{0}^{2\pi}d\phi\, e^{im\phi}{\cal \tilde{H}}\quad,\quad{\cal J}_{m}=\int_{0}^{2\pi}d\phi\, e^{im\phi}{\cal \tilde{P}}\quad,\quad{\cal Q}_{m}=\int_{0}^{2\pi}d\phi\, e^{im\phi}{\cal \tilde{\mathcal{G}}}\,,
\]
the super-BMS$_{3}$ algebra \eqref{eq:Super-BMS-simplest} with central
charges given in (\ref{eq:1}) is recovered.

\section{Reduced super-Liouville-like theory}

In order to obtain the reduced phase space description of the action
\eqref{eq:SuperWZW} on the constraint surface defined by \eqref{constraints},
it is useful to decompose the fields according to 
\begin{eqnarray}
\lambda=e^{\sigma\Gamma_{1}/2}e^{-\varphi\Gamma_{2}/2}e^{\tau\Gamma_{0}}\quad,\quad\alpha=\frac{\eta}{2}\Gamma_{0}+\frac{\theta}{2}\Gamma_{2}+\frac{\zeta}{2}\Gamma_{1}\,,\label{eq:decomposition}
\end{eqnarray}
where $\sigma,\varphi,\tau,\eta,\theta,\zeta$ stand for functions
of $u,\phi$. The constraints \eqref{constraints} then become
\begin{eqnarray}
\sigma^{\prime} & = & e^{\varphi}\,,\nonumber \\
\zeta^{\prime} & = & \mu(e^{\varphi}-\sigma^{\prime})+\frac{1}{2}\sigma^{2}\eta^{\prime}+\sigma\theta^{\prime}\,,\label{eq:constraints_decom}\\
\nu^{- \prime} & = & \frac{1}{\sqrt{2}}\sigma\nu^{+\prime}\,,\nonumber 
\end{eqnarray}
and hence, by virtue of \eqref{eq:decomposition} and \eqref{eq:constraints_decom},
the reduced chiral super-WZW action \eqref{eq:SuperWZW} is given
by
\begin{eqnarray}
I_{R} & = & \frac{k}{4\pi}\int dud\phi\left[\xi^{\prime}\dot{\varphi}-\varphi^{\prime2}+\mu\varphi^{\prime}\dot{\varphi}+\frac{1}{\sqrt{2}}\chi\dot{\chi}\right]\,,\label{IR}
\end{eqnarray}
where $\xi:=-2(\theta+\eta\sigma)+\frac{1}{2}(\nu^-\nu^+)$, and
$\chi:=e^{\varphi/2}\nu^+$. It is worth noting that, in the case
of $\mu=0$, the bosonic part of \eqref{IR} is related to a flat
limit of Liouville theory \cite{Barnich:2013yka}. The super-BMS$_{3}$
generators \eqref{eq:super_BMS_generators} then reduce to 
\begin{equation}
\badat{3}\tilde{\mathcal{H}}=\frac{k}{4\pi}\left(\varphi^{\prime2}-2\varphi^{\prime\prime}\right)\quad,\quad\tilde{\mathcal{P}}=\frac{k}{4\pi}\left(\xi^{\prime}\varphi^{\prime}-\xi^{\prime\prime}+\frac{1}{\sqrt{2}}\chi\chi^{\prime}\right)+\mu\tilde{\mathcal{H}}\quad,\quad\tilde{\mathcal{G}}=2^{1/4}\frac{k}{4\pi\,  }\left(\frac{1}{2}\varphi'\chi-\chi^{\prime}\right)\,,\eadat\label{scgen}
\end{equation}
which generate the following transformations
\begin{eqnarray}
\badat{3} & \delta\varphi=Y\varphi'+Y'\,, \\
 & \delta\xi=2f\varphi'+\xi'Y+2f'-2^{1/4}\ep\chi\,,\label{eq:SLLtransformations}\\
 & \delta\chi=Y\chi^{\prime}+\frac{1}{2}Y^{\prime}\chi+2^{-1/4}\ep\varphi'+2^{3/4}\ep'\,,\eadat
\end{eqnarray}
with $f=T(\phi)+uY'$, $Y=Y(\phi)$, and $\ep=\ep(\phi)$. Therefore, by
construction, the super-Liouville-like theory turns out to be
invariant under \eqref{eq:SLLtransformations}, and the mode expansion
of the algebra of Noether charges is again given by
\eqref{eq:Super-BMS-simplest} and (\ref{eq:1}).

\section{Gauged chiral super-WZW model}

The super-Liouville-like action \eqref{IR}, that has been shown to be
equivalent to the chiral super-WZW model \eqref{eq:SuperWZW} on the
constraint surface given by \eqref{constraints}, can also be described
through a gauged chiral super-WZW model. Here we follow the procedure
given in \cite{Balog:1990mu}, where it was shown that Toda theories
can be written as gauged WZW models based on a Lie group $G$. The
action is endowed with additional terms involving the currents
linearly coupled to some gauge fields that belong to the adjoint
representation of the subgroups of $G$ generated by the step operators
associated to the positive and negative roots.

Hence, we consider the following action principle
\begin{multline}
I[\lambda,\alpha,\nu,A_{\mu},\bar\Psi]=I[\lambda,\alpha,\nu]+\frac{k}{\pi}\int dud\phi\,{\rm Tr}\big[A_{u}(\lambda^{-1}\alpha'\lambda-\frac{1}{4}\lambda^{-1}\nu\bar{\nu}'\lambda-\frac{1}{8}\bar{\nu}'\nu\mathbf{1})\\
+\tilde{A_{u}}(\lambda^{-1}\lambda')-\mu_{M}\tilde{A_{u}}+(\frac{1}{4}\lambda^{-1}\nu')\bar{\Psi}\big]\,,\label{gaugedWZW}
\end{multline}
where $I[\lambda,\alpha,\nu]$ is the flat chiral super-Poincaré WZW
action \eqref{eq:SuperWZW}. Here $A_u, \tilde{A_{u}} $ are along $\Gamma_0$, and
$\mu_{M}:=\mu \Gamma_{1}$ with $\mu$ an arbitrary constant,
while the fermionic gauge field $\bar{\Psi}$ fulfills $[\bar{\Psi}]_{+}=0$
(see Appendix \ref{GWZW} for more details on the construction in
the bosonic case).

One can then show that the action \eqref{gaugedWZW} is invariant (up
to boundary terms) under the transformations given in \eqref{eq:infin_global},
where a subset of the symmetries has been gauged by allowing for an
arbitrary $u$ dependence of the part of
$\sigma,\vartheta$ that belongs to the subspace generated by $\Gamma_{0}$, of
the fermionic parameters that belong to the subspace defined by
$[\bar{\gamma}]_{+}=0$, $[\lambda\gamma]^{-}=0$ and the non-trivial
transformations for the gauge fields are
\begin{eqnarray}
\badat{2} &\, \delta_{\sigma}\tilde{A_u}=-\left(\dot \sigma+[A_u,\sigma]\right)\,,\,  \delta_\gamma\bar{\Psi}=-\partial_{u}\bar{\gamma}\,.\\
& \delta_{\vartheta}A_u=-(\dot \vartheta+[A_u,\vartheta] )\,,\, \delta_{\vartheta}\tilde{A_u}=u\left(\dot \vartheta'+[A_u,\vartheta']\right)-[\tilde{A_u},\vartheta].\eadat
\end{eqnarray}
Therefore, the reduced theory described by the action in \eqref{IR} is
equivalent to the one in \eqref{gaugedWZW}, which corresponds to a WZW
model in which the subgroup generated by the first class constraints
has been gauged. Indeed, the gauge fields $A_u$, $\tilde{A_u}$ and $\Psi$ act as
Lagrange multipliers for these currents, so that the variation of the
action with respect to these non-propagating fields sets them to
zero. In other words, solving the algebraic field equations for the
gauge fields into the action amounts to imposing the first class
constraints \eqref{constraints}, which shows the equivalence of both
descriptions.

\acknowledgments We thank M. Bañados, O. Fuentealba, G. Giribet,
M. Henneaux, P.-H. Lambert, B. Oblak, A. Pérez, and very especially
H. González for enlightening discussions. J.M. and R.T. wish to thank
the Physique théorique et mathématique group of the Université Libre
de Bruxelles, and the International Solvay Institutes for the warm
hospitality.  This work is partially funded by the Fondecyt grants
N${^{\circ}}$ 1130658, 1121031, 3150448. The Centro de Estudios
Científicos (CECs) is funded by the Chilean Government through the
Centers of Excellence Base Financing Program of Conicyt. L.D.~is a
research fellow of the ``Fonds pour la Formation à la Recherche dans
l'Industrie et dans l'Agriculture''-FRIA Belgium. She thanks the group
of the Centro de Estudios Científicos (CECs) for the hospitality.  The
work of G.B.~and L.D.~is partially supported by research grants of the
F.R.S.-FNRS and IISN-Belgium as well as the ``Communauté française de
Belgique - Actions de Recherche Concertees''.

\appendix

\section{Conventions}

\label{conventions}

The orientation has been chosen so that the Levi-Civita symbol fulfills
$\epsilon_{012}=1$, while the tangent space flat metric $\eta_{ab}$,
with $a=0,1,2$, is assumed to be off-diagonal and given by 
\[
\eta_{ab}=\left(\begin{array}{ccc}
0 & 1 & 0\\
1 & 0 & 0\\
0 & 0 & 1
\end{array}\right)\,.
\]
The Dirac matrices in three spacetime dimensions satisfy the Clifford
algebra $\{\Gamma_{a},\Gamma_{b}\}=2\,\eta_{ab}$, and have been chosen
as 
\[
\Gamma_{0}=\sqrt{2}\left(\begin{array}{cc}
0 & 1\\
0 & 0
\end{array}\right)\quad,\quad\Gamma_{1}=\sqrt{2}\left(\begin{array}{cc}
0 & 0\\
1 & 0
\end{array}\right)\quad,\quad\Gamma_{2}=\left(\begin{array}{cc}
1 & 0\\
0 & -1
\end{array}\right)\,.
\]
The matrices fulfill the following useful properties: 
\begin{equation}
\Gamma_{a}\Gamma_{b}=\epsilon_{abc}\Gamma^{c}+\eta_{ab}\mathbf{1}\quad,\quad{{\left(\Gamma^{a}\right)}^{\alpha}}_{\beta}{\left(\Gamma_{a}\right)^{\gamma}}_{\delta}=2\delta_{\delta}^{\alpha}\delta_{\beta}^{\gamma}-\delta_{\beta}^{\alpha}\delta_{\delta}^{\gamma}\,,\label{completeness}
\end{equation}
where $\alpha=+1$, $-1$. The Majorana conjugate is defined as $\bar{\psi}_{\alpha}=C_{\alpha\beta}\psi^{\beta}$,
where 
\begin{equation}
C_{\alpha\beta}=\varepsilon_{\alpha\beta}=C^{\alpha\beta}=\left(\begin{array}{cc}
0 & 1\\
-1 & 0
\end{array}\right)\,,
\end{equation}
stands for the charge conjugation matrix, which satisfies $C^{T}=-C$
and $C\Gamma_{a}C^{-1}=-(\Gamma_{a})^{T}$. Note that this implies
that $\overline{\Lambda^{-1}\psi}=\bar{\psi}\Lambda$, for any $\Lambda\in\mathrm{SL}(2,\mathbb{R})$.
The conjugate of the product of real Grassmann variables is assumed
to fulfill $\left(\theta_{1}\theta_{2}\right)^{*}=\theta_{1}\theta_{2}$.

\section{Gauged chiral bosonic WZW theory}

\label{GWZW}

Let us describe here a way to construct a gauged chiral
$\mathfrak{iso}(2,1)$ WZW model associated to \eqref{eq:SuperWZW} for
the purely bosonic case and $\mu=0$. The action is given by
\begin{eqnarray}
I(\lambda,\alpha)=\frac{k}{\pi}\int dud\phi\,\mathrm{Tr}\left[\dot{\lambda}\lambda^{-1}\alpha'-\frac{1}{2}(\lambda'\lambda^{-1})^{2}\right]\,,\label{flatWZW}
\end{eqnarray}
and it has the following Noether symmetries 
\be
\badat{2}
\delta_{\sigma}\lambda=0 \virg \delta_{\sigma}\alpha=\lambda\sigma(\phi)\lambda^{-1}\,, \\
\delta_{\vartheta}\lambda=-\lambda\vartheta(\phi) \virg \delta_{\vartheta}\alpha=-u\lambda\vartheta'\lambda^{-1}\,.
\eadat
\en
According to \eqref{constraints}, we are interested in gauging the
subset of these symmetries involving the parts of $\sigma$ and
$\vartheta$ along $\Gamma_0$. These parameters are promoted to depend on
both $u$ and $\phi$. \\

One can check that the action
\begin{eqnarray}
I(\lambda,\alpha,A_{\mu})=I(\lambda,\alpha)+\frac{k}{\pi}\int dud\phi\,\mathrm{Tr}\left[-A_{u}\lambda^{-1}\alpha'\lambda+\tilde{A_u}\lambda^{-1}\lambda'\right]\,,
\end{eqnarray}
is invariant under
\be
\badat{2}
&\delta_{\sigma}\lambda=0 \,,\,\delta_{\sigma}\alpha=\lambda\sigma(u,\phi)\lambda^{-1} \,,\, \delta_{\sigma}A_u=0 \,,\, \delta_{\sigma}\tilde{A_u}=-\left(\dot \sigma+[A_u,\sigma]\right),\\
&\delta_{\vartheta}\lambda=-\lambda\vartheta(u,\phi) \,,\, \delta_{\vartheta}\alpha=-u\lambda\vartheta'\lambda^{-1} \,,\, \delta_{\vartheta}A_u=-(\dot \vartheta+[A_u,\vartheta] )\,,\, \delta_{\vartheta}\tilde{A_u}=u\left(\dot \vartheta'+[A_u,\vartheta']\right)-[\tilde{A_u},\vartheta]\,,
\eadat
\en
with $\sigma$ and $\vartheta$ along $\Gamma_0$.

Since the constraints we want to implement set some current components to a
constant, the suitable final action is 
\begin{eqnarray}
I(\lambda,\alpha,A_{\mu})=I(\lambda,\alpha)+\frac{k}{\pi}\int
dud\phi\,\mathrm{Tr}\left[-A_{u}\lambda^{-1}\alpha'\lambda+\tilde{A_{u}}\lambda^{-1}\lambda'-\mu_M
 \tilde{A_{u}}\right]\,,\label{gflatWZW}
\end{eqnarray}
where $\mu_{M}:=\mu\Gamma_{1}$, with $\mu$ an arbitrary
constant, and $A_u, \tilde{A_{u}} $ are along $\Gamma_0$. The action \eqref{gflatWZW}
is indeed still gauge invariant since, as noticed in
\cite{Balog:1990mu}, the variation of $\mathrm{Tr}[\mu_{M}\tilde{A_{u}}]$
under a gauge transformation is a boundary term.

Finally, in order to see how the constraints are explicitly
implemented, it is useful to parametrize the fields according to
\begin{eqnarray}
\lambda=e^{\sigma\Gamma_{1}/2}e^{-\varphi\Gamma_{2}/2}e^{\tau\Gamma_{0}}\quad,\quad\alpha=\frac{\eta}{2}\Gamma_{0}+\frac{\theta}{2}\Gamma_{2}+\frac{\zeta}{2}\Gamma_{1}\,.
\end{eqnarray}
The field equations for the gauge fields imply that
$\sigma'e^{-\varphi}=\mu$ and $\eta'\sigma^{2}+2\theta'\sigma-2\zeta'=0$,
so that, taking $\mu=1$, the reduced action is 
\begin{eqnarray}
I=\frac{k}{4\pi}\int dud\phi\,\left[\xi'\dot{\varphi}-\varphi'^{2}\right]\,,
\end{eqnarray}
where $\xi:=-2(\theta+\eta\sigma)$, in full agreement with the centrally extended BMS$_{3}$ invariant
action found in \cite{Barnich:2013yka}.

\end{document}